\documentstyle[preprint,tighten,aps]{revtex}
\input epsf
\begin{document}
\begin{center}

\Large

{\bf Hydrodynamic lift on bound  vesicles}

\large \vspace{0.5cm}
{\sl  Udo Seifert} \vspace{0.5cm}
\normalsize

 Max-Planck-Institut f\"ur Kolloid- und
Grenzfl\"achenforschung,\\ Kantstrasse 55, 14513 Teltow-Seehof, Germany
 \vspace{0.5cm}
 \end{center}
\begin {abstract} 
Bound vesicles subject to lateral forces such as arising from  shear
 flow are investigated
theoretically by combining a  lubrication analysis of the bound part with
a scaling approach to the global motion.
 A minor inclination of the
bound part  leads to 
significant lift due to the additive effects of lateral and tank-treading
motions. 
 With increasing shear rate, the vesicle unbinds from the substrate at
a critical value. Estimates are in agreement with recent experimental data.
\end{abstract} 
\vspace{0.5cm}
\vspace{0.5cm}
\def\beq{\begin{equation}}
\def\ee{\end{equation}}
\def\pcite{\protect\cite}
\def\pa{\partial}
\def\m{{\bf m}}
\def\q{{\bf q}}
\def\l{{\cal L}^\dagger}
\def\a{{\cal A}}

\def\lsim {\protect
\raisebox{-0.75ex}[-1.5ex]{$\;\stackrel{<}{\sim}\;$}}

\def\gsim {\protect
\raisebox{-0.75ex}[-1.5ex]{$\;\stackrel{>}{\sim}\;$}}

\def\lsimeq {\protect
\raisebox{-0.75ex}[-1.5ex]{$\;\stackrel{<}{\simeq}\;$}}

\def\gsimeq {\protect
\raisebox{-0.75ex}[-1.5ex]{$\;\stackrel{>}{\simeq}\;$}}

\def\u{{\bf u}}
\def\v{{\bf v}}
\def\r{{\bf r}}
\def\ex{{\bf e}_x}
\def\ez{{\bf e}_z}
\def\p{{\partial}}
\def\n{{\bf \nabla}}
\def\gd{{\dot \gamma}}

{\it Introduction.} The {\it equilibrium} aspects of the
interactions between membranes or vesicles and substrates have been
 explored intensely over the last decade \cite{lipo94c}. Quantitative 
experimental data for both the mean shape and the
fluctuations of the bound part of a vesicle
have been obtained by  
using phase contrast microscopy and
reflection interference contrast microscopy  (RICM) \cite{raed94}. 
A qualitative new step concerns the study of bound membranes under
controlled {\it non-equilibrium} conditions such as the behavior of bound vesicles
under shear flow. 
Apart from its fundamental significance, this system can serve as a model
for the biologically ubiquitous situation of adhesion of membranes under
flow. A  prominent example occurs for leucocyte or platelet adhesion
in  capillary flow. Clearly, for biological systems, the non-trivial kinetics
of {\it specific} adhesion molecule pairs under a ramped force 
 contributes essentially to the dynamic
unbinding of these cells (see, e.g., \cite{hamm92,brui95,z:bong95} 
and references therein). 
Still, a thorough understanding of the model case of 
a bound vesicle with 
its interplay between {\it unspecific}
interactions and flow  will be a prerequisite for gaining a  
comprehensive picture of these important dynamic interactions.

A significant experimental step in this direction has  been achieved recently
by combining RICM with a flow
chamber \cite{sims98}. With this set-up one can study the configurations of bound
vesicles under shear flow. It was observed that these vesicles
 unbind from the substrate  at  a
critical shear rate. However, the
effective lift force was found to be about two orders of magnitude larger than
 what was predicted in previous theoretical work \cite{brui95}. 

The purpose of this letter is to analyze theoretically the dynamically induced
interaction between a substrate and  a bound vesicle
under a lateral force such as arising from shear flow.  
This problem is  challenging since it involves two vastly different
length scales. Typically, the vesicle size is of the order of 10 $\mu$m whereas
the distance between substrate and vesicle
is of order 10 nm. A brute force 
approach trying to solve numerically the equations of motion of such a configuration 
as it has been done for  free vesicles in shear flow \cite{z:krau96}
 is bound to require a very fine discretization and, consequently, to face 
high computational costs. A first step in this direction has  been
 achieved recently for the computationally less expensive
 two-dimensional case \cite{dura97,cant98}.

For the experimentally relevant three-dimensional case, a two step approach
will be
followed here. First, the bound part of the vesicle will be treated 
quantitatively within the lubrication approximation which holds if the
 lateral
extension of the bound part is significantly larger than the distance from
the substrate. As a result we  will find that whenever this
bound part is tilted
 a significant hydrodynamic lift arises even for small tilt 
due to the additive effects of translation and relative membrane motion, i.e.
tank-treading. 
In a second step, we couple this lubrication analysis into
a scaling approach of the overall vesicle motion. 
 As a result, we predict a critical lateral force
beyond which vesicles will detach from the bound state and, thus, undergo 
a dynamically induced unbinding transition. This transition must
be distinguished from an equilibrium  unbinding 
transition due either to fluctuations \cite{lipo86,z:seif95g} or a 
competition between
adhesion energy and curvature energy \cite{seif90a}. 

{\it Geometry.} In equilibrium, a vesicle bound to a substrate by a potential
 $V(h)$ acquires
a spherical cap-like shape if the depth $W$ of the potential is 
sufficiently deep, see Fig. 1. The shape can then be characterized  by two
 parameters,
the radius $R$ of the spherical cap and the radius $R_a\leq R/2$ of the
adhesion disc.  The distance $h_0$ of the adhesion disc from the
substrate is determined by the location of the minimum of the adhesion 
potential. In equilibrium, the adhesion disc is parallel to the
substrate but we will here  allow  a small inclination or
tilt angle $\alpha$.

We now apply a force  $F_x$ parallel to the substrate. 
The physical origin of
this force can either be a linear shear field with shear rate $\gd$ or 
a gradient in adhesion energy $\nabla W$ \cite{dura97,cant98}. 
The force then scales as
$F_x \sim \gd \eta R^2$, or $F_x \sim \nabla W R_a^2$, respectively.
 As a result
the vesicle moves with a velocity $v$
into the same direction.
 Since the membrane is fluid, we have
to allow for tank-treading motion which we assume 
for the spherical part to be  a uniform rotation in the
$x,z$ plane at an angular speed
$\bar v_m/R$. This membrane flow on the spherical part enters (or leaves) 
the rim of the adhesion disc
with a $y$ dependent velocity 
\beq
v_m(y) = \bar v_m (1-y^2/R^2)^{1/2},
\label{eq:vm}\ee where $|y|\leq R_a$. The velocities
 $v$ and $\bar v_m$ will later be determined from
force balances but for the moment they are assumed to be given. 
We show first that such a 
 motion generates a significant hydrodynamic lift on the
vesicle.

{\it Lubrication theory.} For future reference, 
the lubrication approximation will be set up not just
for a tilted adhesion disc but 
 for a general membrane
configuration parametrized by $h(\r)$ in a Monge representation
above a substrate at $z=0$ with $\r= (x,y)$. 
The lateral extension $R_a$ of the membrane
is much larger
than the average height $h_0$ above the substrate.
The velocity field is written as
$\v(\r,z) = \u (\r,z) + w(\r,z) \ez$  where $\u(\r,z)$ is the component parallel
to the substrate.
At the membrane, we specify the velocity as
$\v(\r,h(\r)) = \u^h (\r) + w^h(\r) \ez$. 
At the substrate, no-slip boundary conditions imply $\v(\r,0)={\bf 0}$.

The Stokes  equations for the incompressible
fluid between substrate and membrane read
\beq
\eta (\partial^2_z + \n^2)\u = \n p  , \label{eq2}
\ee
and
\beq
\eta (\partial^2_z + \n^2) w = \p_z p \label{eq3} .
\ee
Here $\n$ is the gradient operator in the plane, i.e. $\n=(\p_x,\p_y)$
in Cartesian coordinates. From the continuity equation 
$
\p_z w + \n \u = 0  \label{eq1} ,
$
 it follows that $w/|\u| \sim O(h_0/R_a)$
for small $h_0/R_a$. This scaling is 
the essential observation in lubrication theory.
 It implies via (\ref{eq2}) and (\ref{eq3}) that
the   pressure  is
a function of $\r$ but independent of $z$  to leading order in $h_0/R_a$. Likewise the $\n^2$ term 
can be
ignored compared to the $\p_z^2$ term in (\ref{eq2}).
 The latter equation can hence be
integrated as
\beq
\u(\r,z) =  \n p(\r) \ z(z-h)/2\eta + \u^h(\r) z/h  ,
\ee
which satisfies the boundary conditions at $z=0$ and $z=h$.
 Applying the $\n$ operator to this equation, inserting the resulting
expression into the equation of continuity 
 and integrating the latter over $z$ 
from 0 to $h$
yields a Reynolds-type  equation
\beq
\n^2 p + 3 \n p \n h /h = 12\eta w^h/h^3 + 6 \eta \n( \u^h/h)/h . 
\label{eq4}
\ee

This equation for the pressure holds for any membrane configuration
in the lubrication approximation. We now specialize to the circular
adhesion disc
of radius $R_a$ 
tilted at a small angle $\alpha$, i.e. $h(\r) = h_0 + \alpha x$, see Fig. 1.
No slip
boundary conditions between membrane and fluid
 imply for the translational and tank-treading motion introduced above  
 the boundary values  $w^h(\r) = -\alpha v_m(y)$ and 
$\u^h(\r) = (v-v_m(y)) \ex$ for small $\alpha$. With
$\n h = \alpha \ex$, 
equation (\ref{eq4}) becomes 
\beq
\n^2 p(\r) \approx - 6 \alpha \eta (v_m(y)+v) /h_0^3 
\label{eq:poisson}
\ee  to lowest order in $\alpha$. First, assume that there was
no tank-treading motion, $\bar v_m=0$,  and, hence, no $\r$ dependence of the inhomogeneity in
this 
Poisson equation. The solution then is 
\beq
p(\r) =3\alpha \eta  v (R_a^2-r^2)/ 2h_0^3  + p(R_a)(r/R_a)\cos\phi
\ee
where $p(R_a)\cos\phi$ is the ambient pressure along the rim of the adhesion
disc parametrized by the azimuthal angle $\phi$.  
Integrating the excess pressure over the entire adhesion disc yields the
total lift force on the membrane disc as 
\beq
F_l \equiv \int dA \ p(\r)  = 3 \pi  \alpha \eta v R_a^4 /4 h_0^3 .
\label{eq:lift}
\ee
For $\bar v_m \not = 0$, the solution of the Poisson equation (\ref{eq:poisson})
is slightly
more involved because of the $y$ dependence of the rhs. We are
interested only in the total lift, which involves averaging over the
whole adhesion disc. Given the form (\ref{eq:vm}), it
is then clear that  (\ref{eq:lift}) still holds with $v$ 
replaced by $\bar v_m$ up to a dimensionless
function $f(R_a/R)$ of order unity which
 will not be needed for the scaling analysis to follow.
The important point is that both translational motion and tank-treading motion
contribute similarly to the hydrodynamic lift (\ref{eq:lift}). This lift increases
strongly  with decreasing  distance $h_0$ of the membrane 
from the
substrate. Note that the reversibility of the Stokes equations implies
that there is no hydrodynamic lift for a 
non-tilted configuration with $\alpha =0$.

{\it Scaling analysis of global motion.}
 We now have to link this  lubrication analysis of the adhesion disc to the
overall vesicle dynamics. We apply the force  $F_x$ parallel to the 
substrate 
and assume first  the rotational degree of freedom being locked
at $\alpha =0$.  Then, there are two 
conditions which fix the two velocity parameters $v$ and $\bar v_m$ uniquely. We write these conditions
in scaling form which means that we ignore all numerical prefactors
of the respective terms with the understanding that these prefactors
do not exhibit singular behavior for small  $h_0/R_a$. 

(i) Force balance in the $x$ direction:
\beq F_x \sim \eta \Delta v R_a^2/h_0  + \eta v R  ,
\label{eq:force0}\ee
where $
\Delta v \equiv v - \bar v_m $
is the velocity of the bound part of the vesicle relative to the substrate. 
The first term on the rhs is the lateral  force exerted by
the lubrication layer, the second term is the hydrodynamic drag of the
exterior fluid outside of the lubrication layer.

(ii) Dissipation balance: 
\beq
 F_x v  \sim \eta (\Delta v)^2 R_a^2/h_0  + \eta v^2 R + \eta \bar v_m^2 R .
\label{eq:diss0}\ee 
The lhs represents the power applied by the external force on
the system. The first term on the rhs 
is the dissipation in the lubrication layer. The second one is
dissipation in the exterior fluid outside of the lubrication  layer. 
The third term is the dissipation within the
vesicle due to tank-treading.
We now divide the first equation by $\eta  R$, the second one by 
$\eta v R$ and introduce with 
$
v_S \equiv F_x/6\pi \eta R
$
 the Stokes velocity 
of a spherical vesicle in infinite space.

The two equations (\ref{eq:force0},\ref{eq:diss0}) then read
\beq v_S\sim \beta \Delta v  + v
\label{eq:force}\ee
 and
\beq
v_S \sim \beta (\Delta v)^2/v   +  v + \bar v_m^2 /v  
\label{eq:diss}\ee
where
$
\beta \equiv R_a^2/h_0R $
is a dimensionless variable measuring the relevance of the substrate.
For the scaling analysis, we have to distinguish three cases.

(i) For $\beta << 1$, the vesicle is too far away from the substrate
to be affected significantly in its motion.
We will  not consider  this case further.

(ii) For $\beta \sim 1$, we have $\Delta v \sim v \sim \bar v_m \sim v_S$.
In this case, both translational and tank-treading velocity
 are of the order of the Stokes velocity.

(iii) For $\beta >> 1$, 
equations (\ref{eq:force},\ref{eq:diss})
imply $\Delta v \sim 1/\beta$ and, consequently,
$\bar v_m\approx v \sim v_S$. In this limit, tank-treading and translational velocity become
equal and both are of the order of the Stokes velocity \cite{cant98}. 
 Tanktreading thus  restores (up to factors of order unity) the free mobility
  which would be impossible for
a rigid object with finite $R_a$ so close to a substrate. 
In summary, we can write in each of the two interesting cases 
\beq \bar v_m \sim v \sim v_S \sim F_x/\eta R \ \ \ {\rm and } \ \ \ 
\Delta v/ v \sim h_0R/R_a^2.  \ee

{\it Hydrodynamic lift. }
Using these results, which will hold within perturbation theory also for
 small nonzero tilt angle $\alpha$,  the lift (\ref{eq:lift}) can be written
as
\beq
F_l \sim \alpha \eta v R_a^4/h_0^3 \sim \alpha F_x (R_a/R)(R_a/h_0)^3  .
\label{eq:fl}
\ee 
This expression still depends on the unknown tilt angle $\alpha$ which must be
determined next.

In general, the transversal motion considered so far for the
rotationally locked shape at  $\alpha =0$ generates a torque
 $M$ acting in the $x,z$ plane.
 The origin of this torque are the lateral force and 
hydrodynamic interactions. Their sum will  scale as 
$
M \sim \eta v R^2  \sim F_x R. $
Counteracting to such a torque is a torque arising from the
confining adhesion potential which favors $\alpha =0$. 
The energy  $E(\alpha) $ of a tilted adhesion disc compared to a
non-tilted one 
is given by
$
E(\alpha) \approx (\pi/4) \alpha^2 V'' R_a^4
$
where $V''=V''(h_0)$ is the curvature of the adhesion potential
at the minimum $h_0$. Balancing  the torque $\p_\alpha E$
derived herefrom with  the hydrodynamic one leads to $ 
\eta v R^2 \sim 
\alpha V'' R_a^4  
$ 
or
\beq
\alpha \sim \eta v R^2/V'' R_a^4 \sim F_x R/V''R_a^4 .
\ee
We have written these relations
as if $\alpha$ was positive. Thus, we have implicitly assumed
that the hydrodynamic torque acts to increase $\alpha$. While this cannot be
proven without a full hydrodynamic calculation, evidence for this
assumption in the case of shear flow arises from the fact that for free vesicles
the strain component of the shear orients slightly elongated shapes towards a
positive tilt angle \cite{z:krau96}. Note that if the sign of $\alpha$ was reversed, the
scaling relations derived so far would still hold with lift being replaced by
an additional force pushing the vesicle towards the substrate.

With this expression for the tilt angle $\alpha >0$, we can now
calculate the lift force from (\ref{eq:lift}) as \beq F_l \sim \eta^2
v^2 R^2/h_0^3V'' \sim    F_x^2 /h_0^3 V'' .  \ee For further
evaluation, we  need the specific form of the adhesion potential
$V(h)$. A fairly universal relation can be derived if we assume that
this potential can be  characterized by two scales only, the location
$h_0$ of its minimum and its depth $W\equiv |V(h_0)|$, see Fig. 1.
Then, one has $V''\sim W/h_0^2$, which implies $F_l \sim F_x^2 /W
h_0 .  $

For a comparison with experimental quantities, it is convenient to 
express the adhesion energy $W$ in terms of an effective tension $\Sigma$
using a Young Dupre equation \cite{seif90a}. Except in the tense spherical limit $R_a \to 0$,
both quantities are of the same order, i.e. $\Sigma \sim W$. 
For vesicles in shear flow, we can
thus write for the lift 
\beq
F_l \sim F_x^2 /W h_0 \sim (\eta^2 \gd^2 R^3/\Sigma) (R/h_0).
\ee
This result is a factor of order $R/h_0\simeq 10^2 - 10^3$ larger than a previous theoretical
estimate \cite{brui95}. On the basis of the latter, it was argued in Ref. \cite{sims98}
 that the
experimentally observed lift was 2 to 3 orders of magnitude larger than
theoretically expected. In the light of the present theory, this apparent
discrepancy is most likely due to the factor $R/h_0$ missed in Ref. 
\cite{brui95}.

{\it Dynamical unbinding.}
The lift, if small enough, will displace the 
vesicle slightly from the static equilibrium at $h_0$. The new dynamical
equilibrium position can now  be found  by balancing the lift with the 
restoring force,
$F_z\approx - \pi R_a^2 (h-h_0) V'', $ arising from an expansion of the
potential around its minimum. The relative shift thus becomes
\beq
(h-h_0)/h_0 \sim F_x^2/h_0^4 V''^2 R_a^2 \sim F_x^2/W^2R_a^2 .
\ee

Typically, if $(h-h_0)/h_0) \sim 1$, the lift will be too strong to be compensated by
the attractive potential. Then, 
 the vesicle will unbind from the substrate under
the action of a lateral force. Using this criterion, we can determine the
critical lateral force $F_x^c$ as
$
F_x^c \sim h_0^2 V'' R_a \sim WR_a.
\label{eq:fxc}
$
If the lateral force arises from a shear field,
the critical shear rate is
\beq
\gd^c \sim WR_a/\eta R^2.
\label{eq:gc}\ee

Depending on the specific conditions, measurements for the
 adhesion energy $W$
have obtained
a vast range of values from 10$^{-1}$ to 10$^{-6}$ erg/cm$^2$ 
\cite{evan87,serv89,raed94}. Consequently, the
critical shear rate will also depend strongly on the conditions. Rather than
using (\ref{eq:gc}) to estimate this force, we chose a typical shear rate of
 $\gd = 1/$s 
and determine the critical adhesion energy $W^c$ at which
 the dynamical unbinding should occur.
For a vesicle with  $R = 10\mu$m, $R_a/R = 0.1$
and a fluid viscosity of $\eta = 10^{-2}$erg s/cm$^3$, we find a
critical adhesion energy of $W^c \sim 10^{-4}$erg/cm$^2$ which
is well within the above range. For $W>W^c$, vesicles with the
above geometry will unbind from the
substrate. 

The  tilt angle $\alpha^c$ at the critical value
is 
\beq
\alpha^c   \sim F_x^c ~ R/V''R_a^4 \sim F_x^c ~R h_0^2/WR_a^4
 \sim h_0^2 ~R/R_a^3.
\ee
Within this simple assumption about the adhesion potential, the
scaling of the critical tilt is thus predicted to be
 determined exclusively by geometrical quantities. Note that
since  $h_0<< R_a\lsim R$, the critical tilt angle can be
very small, e. g., $\alpha_c \simeq 10^{-3}$ for $h_0=10$ nm 
and $R,R_a$ as
above.

{\it Beyond unbinding.}
What happens with the vesicle for larger lateral forces
$F_x>F_x^c$~? If the
adhesion potential decays to 0 for large $h$, one can expect that the
vesicle will continously drift away from the substrate presumably in a combination of
tank-treading and tumbling motion. This regime has been studied perturbatively
for large
$h_0/R$ in Ref.\cite{z:olla97a}.

If, however, as often in experiments, 
the vesicle is filled with a slightly denser fluid than the
surroundings, gravity will keep it close to the substrate
since the lift necessarily gets weaker with increasing distance. 
Such a state where after detachment
the vesicle still keeps its spherical cap like configuration
and  translates at about 100 nm above the substrate
  has been reported experimentally \cite{sims98}. 
We close with a somewhat speculative theoretical analysis
   of this state.
The jump in separation from the substrate of about 100 nm
could indicate that a short range adhesion potential is no longer responsible to the
energetics after detachment. 
Neither can the potential then exert a restoring torque. Therefore, the
total hydrodynamic torque  must vanish. 
At which angle $\alpha$ this 
happens (if at all) can only be determined by a full  hydrodynamic
theory. The present scaling approach, however, allows to relate
this  tilt angle $\alpha$ to the  separation $h$ by balancing
the lift force (\ref{eq:fl}) with the gravitational force $F_g = 
 g \Delta \rho V$, where $g\simeq 1000$ cm/s$^2$ and $\Delta \rho
$ is the
density difference. Such a balance leads to 
$
\alpha \sim (h_0/R_a)^3 ( F_g/ F_x)(R/R_a).
 $
 Using typical experimental values
$F_g\simeq10^{-7}$erg/cm, $F_x \simeq 10^{-8}$erg/cm, $h_0/R_a\simeq 0.1$,
and
  $R/R_a \simeq10$,  one finds with $\alpha \simeq 0.1$ 
that a quite small tilt angle
could generate enough lift to sustain such  a stationary state \cite{note1}.
It must be 
emphasized that this estimate  is  speculative in two respects. First,
we have not shown that the full hydrodynamics allows such a torqueless
 state but have
rather postulated it. Secondly, numerical prefactors could easily 
conspire and change such an estimate about one order of magnitude.  

{\it Summarizing perspective.}
We have analysed the hydrodynamics of a bound
vesicle under a lateral force allowing for tank-treading motion. A positive
tilt of the bound part will generate lift. 
By combining a lubrication analysis 
with scaling arguments, a dynamically induced unbinding transition
 is predicted when the 
lateral force exceeds a critical value. Even though our analysis used
for simplicity the spherical cap configuration, the
scaling relations will also hold for weakly bound states with
rounded
contact regions \cite{seif90a}. More quantitative theoretical calculations
 will be needed to fill
in  numerical prefactors necessarily missing in such a
scaling analysis. Such full scale calculations will also reveal the 
role of transients and assess whether the theoretically
somewhat speculative 
stationary translating
state after unbinding exists. On the experimental side,
special efforts should now be undertaken to measure the here 
theoretically
predicted but
so far not yet experimentally reported tilt angle  which, 
even if minute,  is
ultimately the source of the lift.

{\it Acknowledgments.} I thank R. Lipowsky for pointing my attention to Ref. 
\cite{sims98}; 
 E. Sackmann, R. Simson and H.G. D\"obereiner for stimulating discussions;
I. Cantat and C. Misbah for the same and for 
sending Ref. \cite{cant98} prior to publication.


\begin{thebibliography}{10}
\bibitem{note1}
Note that this estimate is larger than the critical value derived above for
  unbinding from a short-range potential.
\end{thebibliography}

\begin{figure}[b]
\epsfxsize=12cm
\begin{center}
\leavevmode
\epsfbox{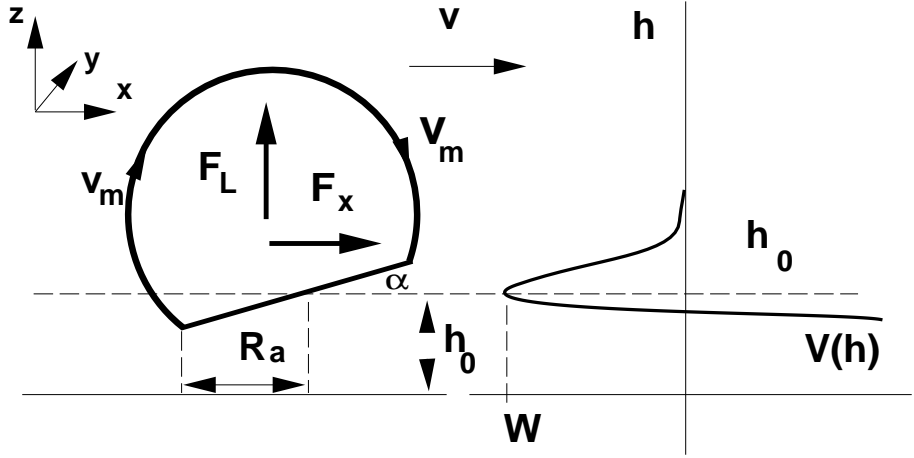}
\end{center}
\caption{A bound vesicle under a lateral force $F_x$ (left)
in a potential well $V(h)$ with minimum at $h_0$ and depth $W$ (right). 
The radius  of the adhesion disc is $R_a$,
its tilt angle is $\alpha$. The vesicle translates at velocity $v$ and
tank-treads at velocity $v_m$. This motion generates a lift $F_l$.}
\end{figure}

\end{document}